\documentclass[prl,aps,floats,amsmath,amssymb,twocolumn,superscriptaddress,preprintnumbers,nofootinbib]{revtex4}

\usepackage{epsfig,latexsym,cancel,amssymb,amsmath}
\usepackage{epstopdf}
\usepackage{color}
\usepackage{dcolumn}
\usepackage{bm}
\usepackage{hyperref}
\usepackage{slashed}

\usepackage{graphicx}

\def\be{\begin{equation}}
\def\ee{\end{equation}}
\def\bea{\begin{eqnarray}}
\def\eea{\end{eqnarray}}

\def\bbuildrel#1_#2^#3{\mathrel{\mathop{\kern 0pt#1}\limits_{#2}^{#3}}}
\def\slash#1{\setbox0=\hbox{$#1$}#1\hskip-\wd0\dimen0=5pt\advance
       \dimen0 by-\ht0\advance\dimen0 by\dp0\lower0.5\dimen0\hbox
         to\wd0{\hss\sl/\/\hss}}

\newcommand{\gae}{\lower 2pt \hbox{$\, \buildrel {\scriptstyle >}\over {\scriptstyle
\sim}\,$}}
\newcommand{\lae}{\lower 2pt \hbox{$\, \buildrel {\scriptstyle <}\over {\scriptstyle
\sim}\,$}}

\newcommand{\beq}{\begin{eqnarray}}
\newcommand{\eeq}{\end{eqnarray}}

\newcommand{\ba}{\begin{array}}
\newcommand{\ea}{\end{array}}

\long\def\symbolfootnote[#1]#2{\begingroup%
\def\thefootnote{\fnsymbol{footnote}}\footnote[#1]{#2}\endgroup}

\def\lsim{\mathrel{\rlap{\lower4pt\hbox{\hskip1pt$\sim$}}
    \raise1pt\hbox{$<$}}}         
\def\gsim{\mathrel{\rlap{\lower4pt\hbox{\hskip1pt$\sim$}}
    \raise1pt\hbox{$>$}}}         

\def\lsim{\:\raisebox{-0.5ex}{$\stackrel{\textstyle<}{\sim}$}\:}
\def\gsim{\:\raisebox{-0.5ex}{$\stackrel{\textstyle>}{\sim}$}\:}

\def\beq{\begin{equation}}
\def\eeq{\end{equation}}
\def\bea{\begin{eqnarray}}
\def\eea{\end{eqnarray}}
\def\to{\rightarrow}

\begin{document}
\DeclareGraphicsExtensions{.jpg,.pdf,.mps,.png,}

\title{
Multiple $b$-jets Reveal Top Super-partners and the 125 GeV Higgs
}

\author{David Berenstein}
\affiliation{Department of Physics, University of California,
Santa Barbara, CA 93106, USA}

\author{Tao Liu}
\affiliation{Department of Physics, University of California,
Santa Barbara, CA 93106, USA}

\author{Erik Perkins}
\affiliation{Department of Physics, University of California,
Santa Barbara, CA 93106, USA}

\begin{abstract}
We demonstrate that in supersymmetry (SUSY) with relatively light top superpartners, $h\to b\bar b$ can be a very promising channel to discover the SM-like Higgs resonance at the Large Hadron Collider (LHC), although in general contexts it is thought to be challenging because of its huge QCD background. In this scenario, the SM-like Higgs boson is mainly produced via cascade decays initiated by pair-produced stop or sbottom squarks. The good sensitivity to $h\to b\bar b$ owes a great deal to the application of multiple ($\ge 4$) $b$-jet tagging in removing the QCD background, and color-flow variables for reconstructing the Higgs resonance. We show in two benchmark points that a SM-like Higgs resonance can be discovered at the 14 TeV LHC (with a signal-to-background ratio as high as $0.35$), with $\sim 40$ fb$^{-1}$ of data. Potentially, this strategy can be also applied to non-SUSY theories with cascade decays of top partners for the SM-like Higgs search, such as little Higgs, composite Higgs, and Randall-Sundrum models.
\end{abstract}

\maketitle


Deciphering the Higgs mechanism is one of the top priorities of the ATLAS and CMS experiments at the Large Hadron Collider (LHC). Recently, both the CMS and the ATLAS collaborations announced discovery of a Higgs-like resonance, based on a combined analysis for the Standard Model (SM) Higgs searches via $h\to \gamma\gamma$ and $h\to ZZ^* \to 4l$, with the  reconstructed invariant mass $\sim 125-126$ GeV~\cite{:2012gu,:2012gk}. Furthering the identification of this new particle greatly enhances the necessity and emergency of information from complementary Higgs search channels like $h\to b\bar b$.

Unlike $h\to \gamma\gamma$ and $h\to ZZ^* \to 4l$, $h\to b\bar b$ can help test the mass generation mechanism of the SM fermions directly. However, the $h \to b \bar b$ search at the LHC is  challenging because of its huge QCD background, except in the highly boosted regime where jet kinematics allows for the successful application of jet-substructure tools~\cite{Butterworth:2008iy,Plehn:2009rk}. Physics beyond the SM can modify the collider phenomenology of the Higgs search modes if there are new Higgs production mechanisms available. With new strategies designed, the sensitivity of measuring the challenging decay modes could be sizably improved. In this letter we will demonstrate that in supersymmetric scenarios with relatively light top partners, the Higgs discovery via $h\to b\bar b$ could be greatly assisted by a set of dedicated strategies.

Supersymmetry is the prime candidate theory for physics beyond the SM. It provides a natural solution to the hierarchy problem. Among various SUSY scenarios, the ones with relatively light top partners like natural SUSY are theoretically more predictive and experimentally more accessible.  For example, in natural SUSY it is predicted that there existed light superparticles~\cite{Papucci:2011wy}: two stop and one left-handed sbottom squarks, with their masses $\lesssim 700$ GeV; two Higgsino-like neutralinos and one Higgsino-like chargino, with their masses $\lesssim 350$ GeV; and a gluino, with its mass $\lesssim 1500$ GeV. (In this letter, however, we do not adhere to a precise measurement of naturalness, but focus on general scenarios with relatively light top partners.)

\begin{figure}[!h]
\centering
\includegraphics[width=0.23\textwidth]{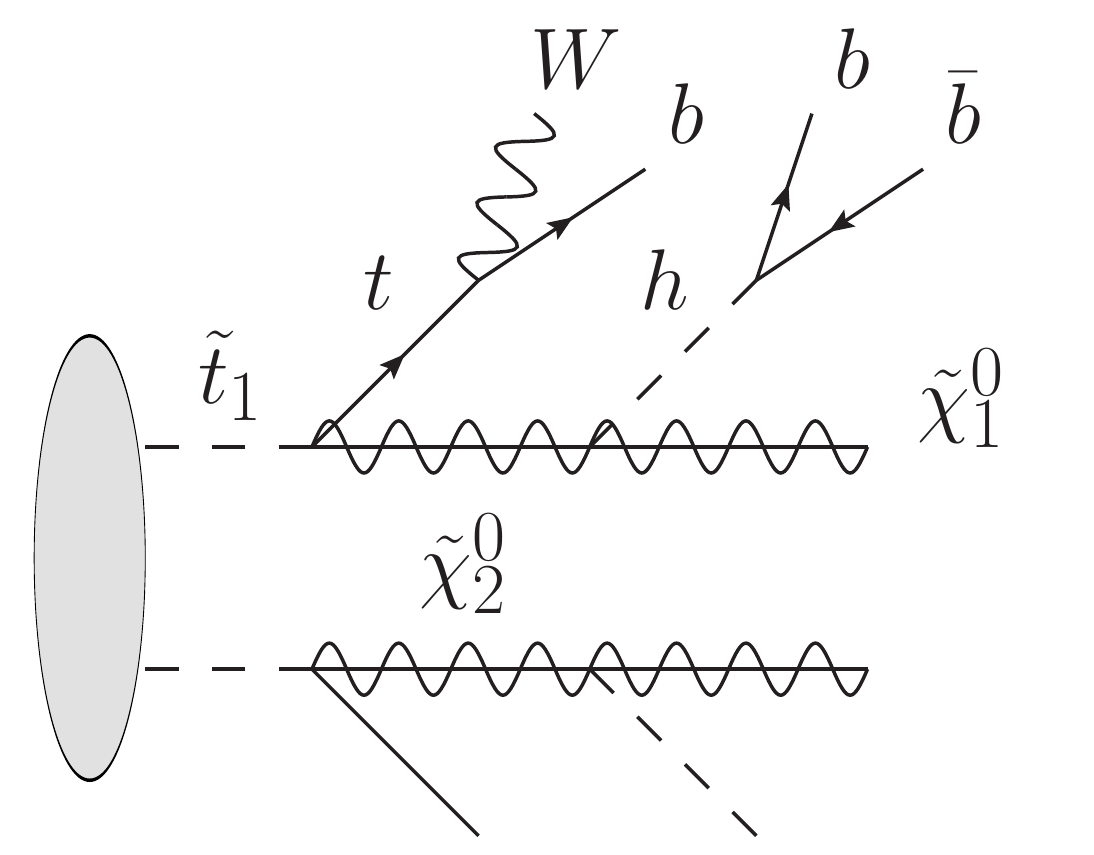}
\includegraphics[width=0.23\textwidth]{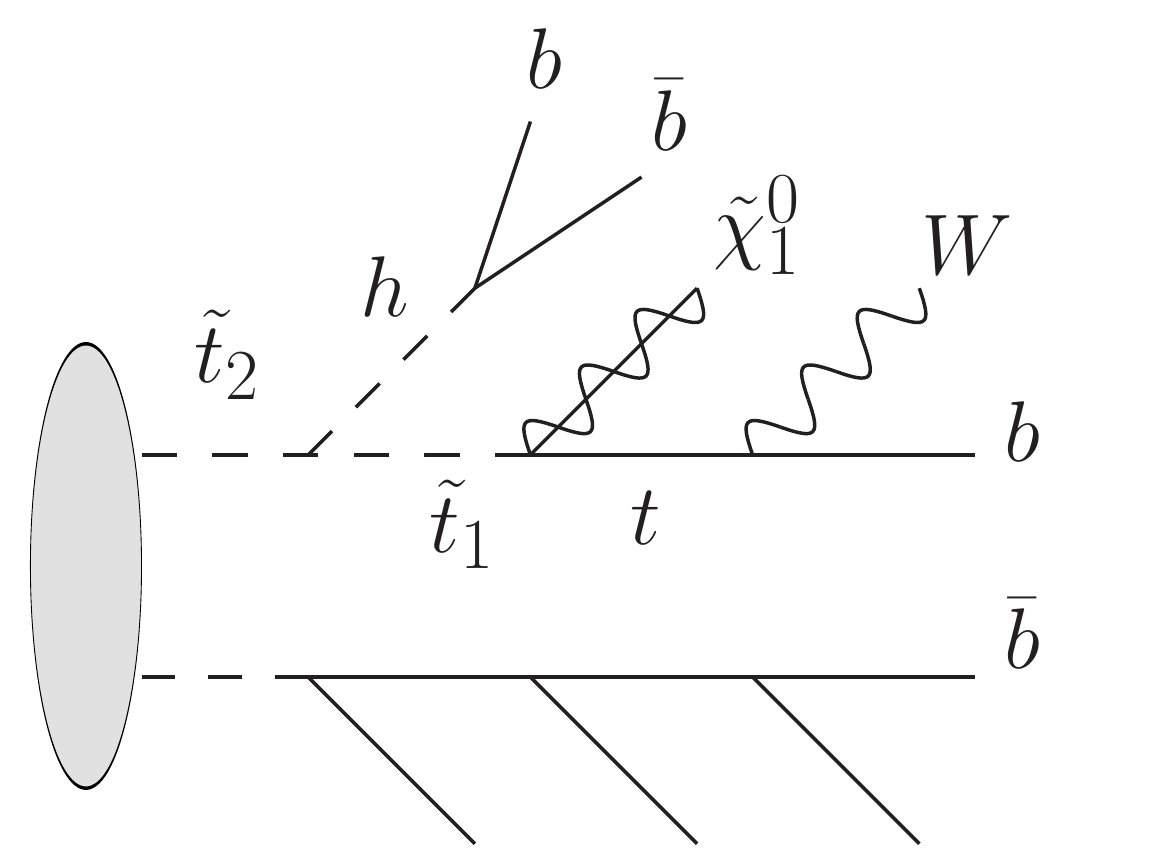}
\caption{The primary Higgs-producing SUSY cascades in the first (left) and second (right) benchmark scenarios.}
\label{mec}
\end{figure}

With conserved R-parity, stop and sbottom squarks are pair-produced at colliders. An interesting feature is that they often decay into a SM-like Higgs boson, in association with a $b$ quark (for discussions on SUSY-assisted Higgs production in various contexts, see~\cite{Baer:1992ef}-\cite{Kribs:2009yh}).  If the Higgs boson decays into a pair of $b$ quarks, the final state typically  contains at least four $b$ quarks. There are two main mechanisms for the Higgs production via the stop and sbottom cascade decays. In the first case, the Higgs boson is produced via neutralino or chargino decays where the Yukawa couplings of the SM fermions are not directly involved.
In the second case, the Higgs boson is directly produced via the decay of the heavier stop squark where the top Yukawa coupling gets involved. Because these processes are initiated by light squarks, the produced Higgs bosons tend to be less boosted, compared to the case discussed in~\cite{Kribs:2009yh} (for discussions on less boosted Higgs boson, also see~\cite{Gori:2011hj}). Two topologies of these mechanisms are illustrated in Fig.~\ref{mec}. Given that the QCD events containing multiple $b$-jets are relatively few, this feature provides a new strategy of improving the sensitivity of the $h \to b\bar b$ resonance search at the LHC.

To design the minimal strategies that ensure optimal coverage of the full space of models, we perform a collider analysis in two benchmark scenarios which represent these two mechanisms, respectively. Their mass spectra and branching ratios are calculated using SUSY-HIT~ \cite{Djouadi:2006bz}, in the context of the MSSM with $m_h=125$ GeV set as an input (see Table~\ref{T:BenchmarkPoints}). Here new contributions to $m_h$ have been implicitly assumed which could be either from new F-terms or from new D-terms (for recent discussions, e.g., see~\cite{Hall:2011aa}). These benchmarks are not fully realistic, but simplified models in essence~\cite{ArkaniHamed:2007fw}. Generalization of the analysis to more realistic cases is straightforward. Though searching for fully realistic benchmark points is not our main motivation, we would like to point out that these benchmark points are still safe and are allowed by current experimental bounds, after comparing them with the publicly available searches, such as the CMS $b$-enriched razor-variable search~\cite{:CMS-PAS-SUS-11-024}, the ATLAS light sbottom search~\cite{:ATLAS-CONF-2012-106}, and the CMS $b$-jets $+ \slash{E}_T$ search~\cite{:2012rg}.

\begin{table}[!h]
\centering
\begin{tabular}{|c|c|c|}
\hline
Benchmarks & I (GeV or \%) & II (GeV or \%)\\
\hline
$m_{\tilde g}$           & 1281 & 1264 \\
$m_{\tilde t_1} $        & 568  & 260  \\
$m_{\tilde t_2} $        & 682  & 586  \\
$m_{\tilde b_1}$         & 567  & 555  \\
$m_{\tilde \chi^0_1}$    & 87   & 84   \\
$m_{\tilde \chi^0_2}$    & 325  & 415  \\
$m_{\tilde \chi^0_3}$    & 336  & 433  \\
$m_{\tilde \chi^\pm_1}$  & 321  & 413  \\
$m_{h}$                  & 125 & 125   \\
\hline
${\rm Br}(\tilde t_2 \to \tilde t_1 + h)$          & 0  & 47 \\
${\rm Br} (\tilde t_1 \to \tilde \chi_1 + h + t )$ & 52 & 0  \\
${\rm Br}(\tilde h \to b\bar b )$                  & 61 & 61 \\
\hline
\end{tabular}
\caption{Two benchmark scenarios. The weak-scale mass spectrum and decay branching ratios are calculated using SUSY-HIT~ \cite{Djouadi:2006bz}, with $m_h=125$ GeV set as an input.}\label{T:BenchmarkPoints}
\end{table}

At the LHC, the resonance search of the SM-like Higgs boson via multiple $b$-jets has three main backgrounds. The first one is the SM background, mainly $t\bar{t}b\bar{b}$. The $t\bar{t} + \leq 2j$ is potentially important, but the 4$b$-tagging requirement can remove much of this background; this is discussed further below. Other SM processes involving $\geq 4b$ jets in the final state are less important for our purposes; \emph{e.g.} we have checked that contributions from QCD $b b \bar b \bar b$ can be efficiently removed with a $\slashed{E}_T$ and $H_T$ requirement, and that $ZZ\rightarrow 4b$ makes a negligible contribution to our search. The second type of background is SUSY events containing multiple $b$-jets, but with no $h$ produced. The third background is combinatorial, arising from events where a SM-like Higgs boson decaying to $b$ quarks is present in the cascade, but where the $b$-tagged jets paired to reconstruct the Higgs are chosen incorrectly. We stress, however, that the latter two types are background \emph{for the Higgs reconstruction only} - in fact these `backgrounds' constitute a strong signal for SUSY.

\begin{figure}[!h]
\centering
\includegraphics[width=0.3\textwidth]{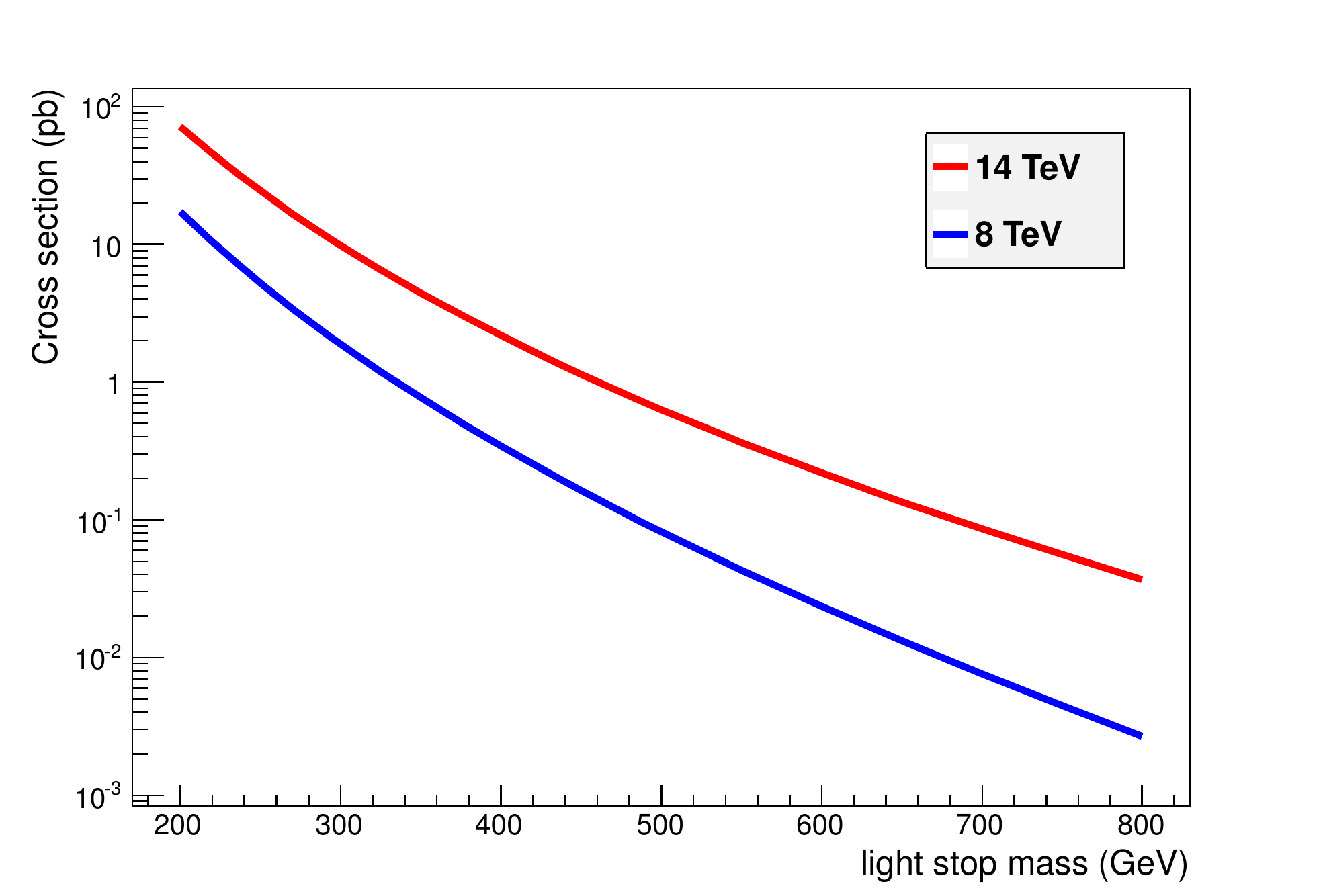}
\caption{The NLO cross section of stop pair production at the $8$ TeV and $14$ TeV LHC generated by Prospino2 ~\cite{Beenakker:1997ut}. }
\label{stop_cross_section}
\end{figure}

The cross sections for all processes here include a $K$ factor; \verb+Prospino2+~\cite{Beenakker:1997ut} determined this for the SUSY processes. The $K$ factor for pure QCD $t\bar{t}b\bar{b}$ and $t\bar{t}+$jets is 2.3, while that for contributions to $t\bar{t}b\bar{b}$ from $t\bar{t}Z$ and $t\bar{t}h$ is 1.6~\cite{Bredenstein:2009aj}.
The next-to-leading-order (NLO) cross section of stop pair production at the $8$ TeV and $14$ TeV LHC is shown in Fig.~\ref{stop_cross_section}.

Our analysis framework is as follows. We generate events with \verb+MadGraph5 / MadEvent+~\cite{Alwall:2011uj}, and perform showering and hadronization with \verb+Pythia+ 8.1~\cite{Sjostrand:2007gs}. For the SM backgrounds, we use the \verb+Pythia-PGS+ package in \verb+MadGraph+ to do showering and MLM matching. We do not employ any dedicated detector simulation in our analysis, but we do place cuts on the hadron-level kinematics to mimic the response of a generic detector. We subject the 4-momentum of each visible final state particle to a Gaussian smearing, as implemented in \verb+Delphes 2.0+~\cite{Ovyn:2009tx}. After smearing, all visible final state particles are required to have $p_T > 0.9$ GeV, and $|\eta| < 5.0$ to fall within the calorimeter acceptance. The remaining `tracks' are identified using generator-level information.

Jet reconstruction is performed next, using \verb+FastJet 3.0+~\cite{Cacciari:2011ma}. We use the inclusive anti-$k_T$ algorithm with $R = 0.5$ and $p_T > 20$ GeV, performing the clustering using all tracks. This sometimes results in `jets' which are actually isolated leptons and photons. Electrons and photons are identified as isolated if the scalar sum of charged track $p_T$ in a cone of $R = 0.2$ does not exceed 10\% of the electron or photon $p_T$; we further require that the electrons have $p_T > 20$ GeV and $|\eta| < 2.47$, while isolated muons are instead required to have $p_T > 10$ GeV and $|\eta| < 2.4$, with the scalar $p_T$ sum of charged tracks within $R = 0.2$ of the muon less than 1.8 GeV~\cite{Ovyn:2009tx}. Once isolated leptons and photons are identified, they are removed from the collection of clustered objects. The remaining jets are then flavor-tagged with generator-level information.

B tagging is done by first determining the parton with highest $p_T$ within the jet cone, and considering this to be the `Monte-Carlo-true' flavor of the jet. True $b$-jets are tagged with an efficiency parameterized by $\varepsilon = 0.6\tanh(p_T/36.0)(1.02-0.02|\eta|)$~\cite{ATLAS-CONF-2011-089}, and true c-jets and light jets are mistagged as $b$-jets at flat 10\% and  1\% rates~\cite{Aad:2009wy,CMSTDR}, respectively. Jets are only considered for $b$-tagging if they fall within the tracker acceptance of $|\eta| < 2.5$. Finally, each final-state object is assigned an ancestor heavy particle according to the generator-level decay history; jet ancestry is determined by the ancestry of the hardest parton in the jet cone.

After the final state objects have been reconstructed, events are subjected to selection cuts. Our analysis implements no triggering step, but it is useful to consider this issue briefly. The current searches at ATLAS~\cite{:ATLAS-CONF-2012-135} and CMS~\cite{:CMS-PAS-HIG-12-025} for $t \bar{t} h$ (which has similar final states and backgrounds to our scenarios) use a low-level trigger based on isolated leptons. However, such triggers do not have a high acceptance for our benchmark points. The focus of our strategy is on heavy flavor; it is possible to perform $b$-tagging at the trigger level, so one might consider a trigger on multiple $b$-tags. Actually, at 8 TeV b-tagging is already used at the trigger level, although more as a last resort~\cite{TomDanielson}. We would highly suggest explicit multi-object triggers with b-tagging at 14 TeV. Alternatively, simpler $H_T$ and $\slashed{E}_T$ triggers help capture the multiple-jet and SUSY nature of our events without the need for intensive online computation. $H_T > 500$ GeV has a high efficiency for our benchmark points at 14 TeV; $\slashed{E}_T$ has a lower efficiency, but a modest requirement can help reduce much of the QCD background. As a comparison, at 8 TeV hadronic $p_T$ sum (300 GeV) + MET requirement (100 GeV) is assumed in the CMS experiment~\cite{CMS:trigger}. But, pileup potentially may make these simple triggers less robust.

\begin{figure}[h]
\centering
\includegraphics[width=0.3\textwidth]{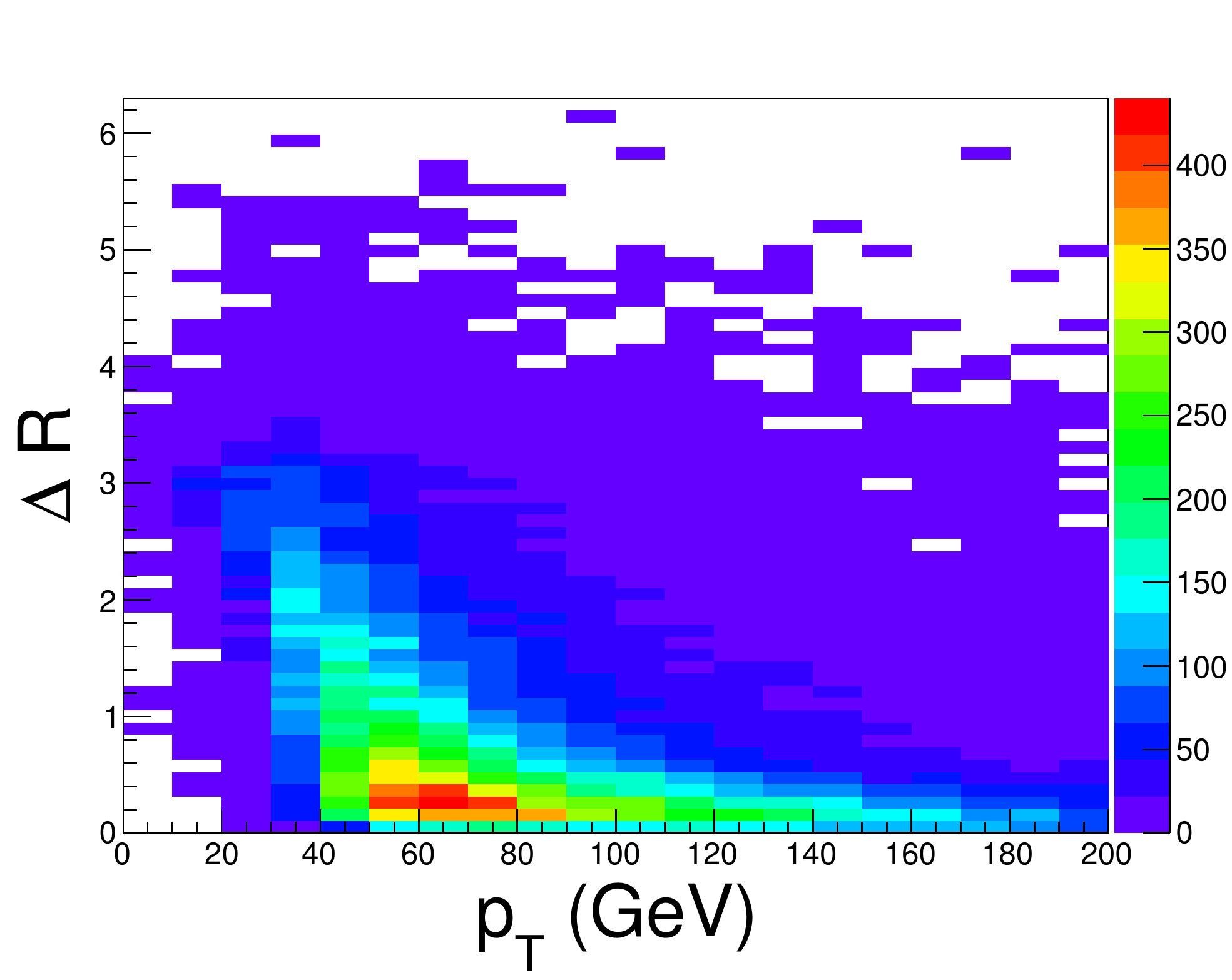}
\caption{$p_T$ vs. $\Delta R$ for non-top $b\bar{b}$ quark pairs in $t\bar{t}b\bar{b}$ events.} \label{Gsplit}
\end{figure}

For the first benchmark point, we apply the following cuts at 14 TeV:
(1) at least 6 jets with $|\eta| < 2.8$,
(2) at least 4 $b$ tagged jets, at least one with $p_{T} > 30$ GeV,
(3) $\slashed E _T > 150$ GeV,
and (4) $H_T > 500$ GeV.
For the second benchmark point, we amend the last two cuts to (3) $\slashed E _T > 120$ GeV, and (4) $H_T > 650$ GeV. In addition to the QCD multiple-jets,  the $4b$-tag requirement can also efficiently remove the $t\bar t b\bar b$ background. The two non-top $b$ quarks in $t\bar t b\bar b$ events are mainly generated by gluon-splitting. This induces these events to fail the $4b$-tag cut for two reasons. The first one is at the level of MC truth. The $b$ quarks from gluon splitting tend to be softer than others, and so recoil more dramatically during parton showering. This makes the hard-process $b$-quarks less aligned with the resulting jet, leading to a larger failure rate in parton-jet flavor matching. In addition, $b$-quarks from gluon splitting tend to be more collimated, the collimation increasing with the $p_T$ of the $b\bar{b}$ pair; this can be seen at parton level in Figure~\ref{Gsplit}. This effect causes the resulting jets to overlap and be reconstructed and tagged as a single $b$ jet.  These effects are implicitly indicated in the cut flows of Table~\ref{cutflow}.

Next, we reconstruct the Higgs resonance from the $b$-jets in the remaining events. At this stage, we encounter a combinatorial problem. Given that this is a dedicated resonance search, we define the signal sample as the correctly reconstructed Higgs bosons or the correctly selected $b$-jet pairs. If a wrong $b$-jet pair is selected as the Higgs candidate for a Higgs event, even if its invariant mass falls into the yet-to-be-known Higgs mass bins, we will count it as ``combinatorial background" instead of a particle signal. This is different from the definition of the signal sample in the studies on the SM Higgs search via the $tth$ process~\cite{Aad:2009wy,:ATLAS-CONF-2012-135} and the SUSY non-standard Higgs search via the processes associated with additional $b$-jets~\cite{Dai:1994vu}, where the Higgs event with a wrongly selected $b$-jet pair as the Higgs candidate, if the $b$-jet pair has a ``correct" invariant mass, was counted as signal instead of background.

The local significance of a resonance is simplest to interpret when only one dijet per event lies in the local mass window $\Delta m_h$, but this method effectively may result in the loss of some correct $b$-pairs. Multiple pairs can be chosen per event as long as their invariant masses differ by more than $\Delta m_h$. To achieve this, we rank the jet pairs according to various variables, always keeping the first-ranked pair. The second-ranked pair is included if $|m_1 - m_2| > \Delta m_h$, and the $n^{th}$-ranked pair is included if $|m_i - m_n| > \Delta m_h$ with $i$ running over all $b$-jet pairs included before. $\Delta m_h$ is mainly controlled by the resolution for jet reconstruction at the LHC - in our analysis, $\Delta m_h = 40$ GeV is assumed. Though an event may contribute more than one pair of $b$-jets for the resonance reconstruction, within the mass window all $b$-jet pairs are from different events.

\begin{figure}[h]
\centering
\includegraphics[width=0.23\textwidth]{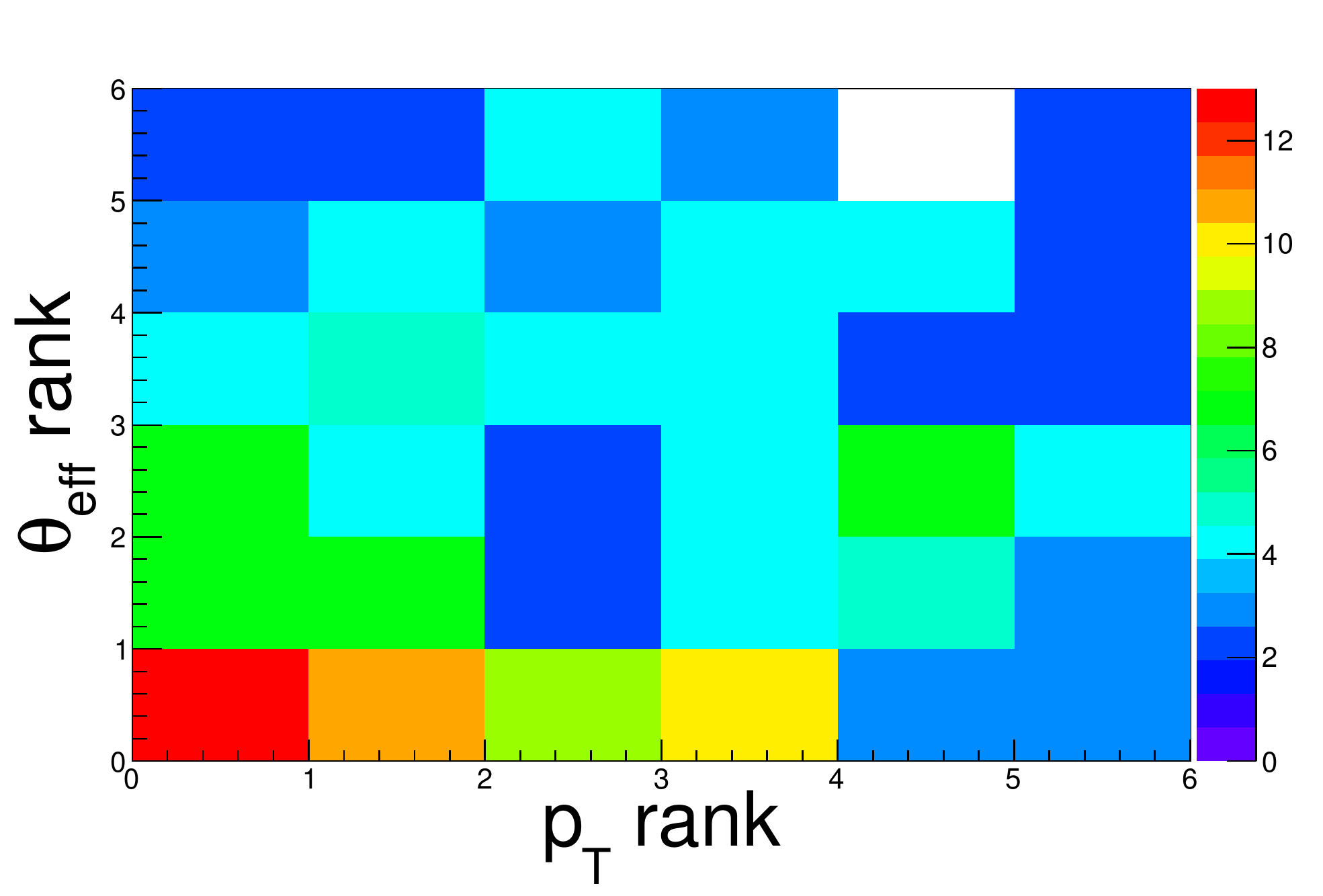}
\includegraphics[width=0.23\textwidth]{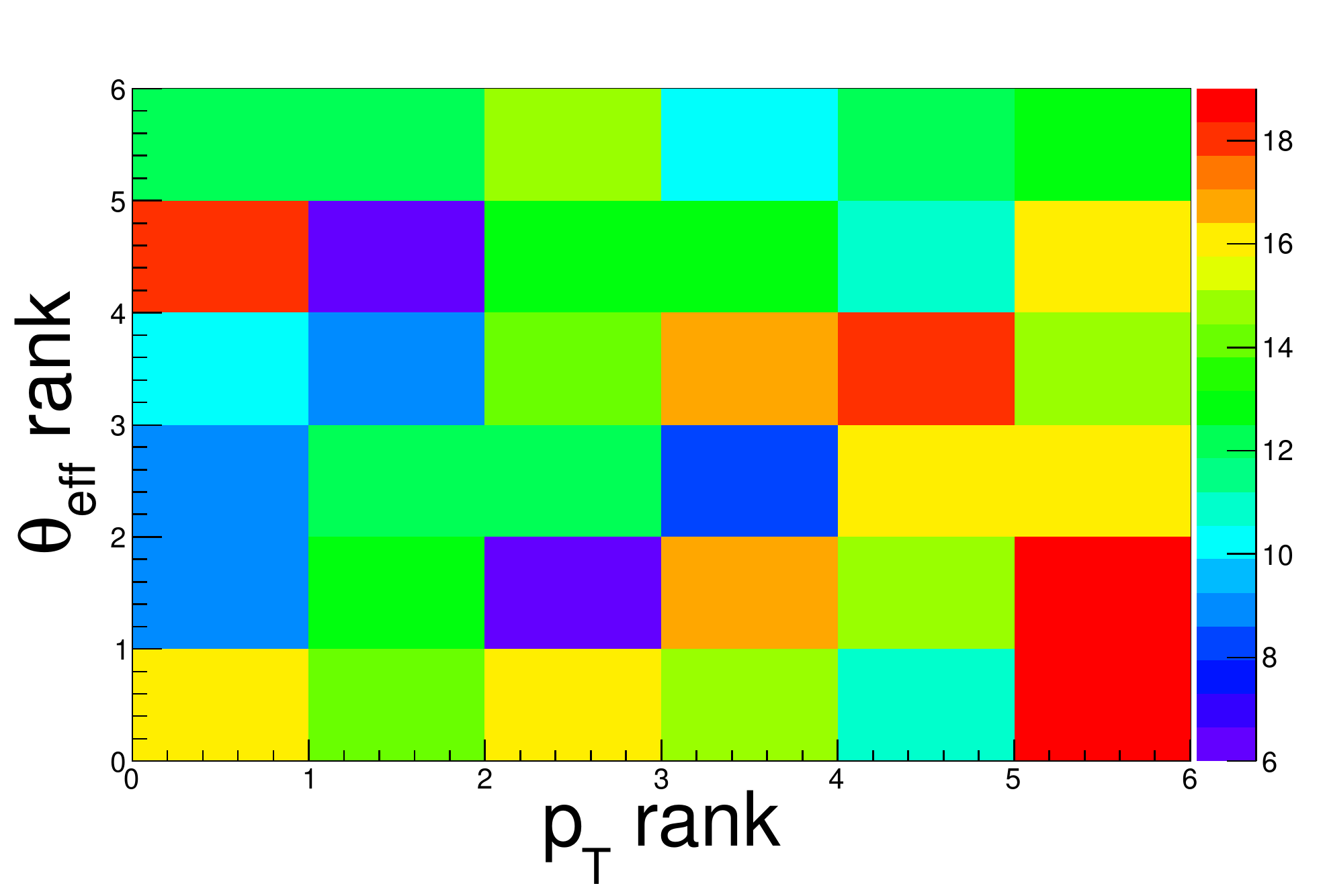}
\caption{$p_T$ vs. $\theta_{\rm eff}$ ranking for correct (left) and wrong (right) $b$-jet pairs falling inside the mass window $[100, 140]$ GeV in signal events, Case I at 14 TeV.} \label{rank}
\end{figure}

To rank the $b$-jet pairs in each event, the kinematics of jet pairs and jet superstructure~\cite{Gallicchio:2010sw} are employed. We use the $p_T$ of each $b$-jet pair, as well as the `pull angle' of the pair. The `pull' of a jet is the vector in the $y-\varphi$ plane defined by
$\vec{t} = \sum_{i\in\textrm{jet}} \frac{p_T^i |r_i|}{p_T^{\textrm{jet}}}\vec{r}_i$
where $\vec{r}_i = (\Delta y_i, \Delta\varphi_i)$ is the displacement of the $i^{th}$ jet component from the jet axis. $\vec{t}$ is a measure of the hadronic energy gradient within the jet, and carries information about how the jet's ancestral parton hadronized. In particular, a pair of jets originating from quarks pair-produced by a color singlet tend to hadronize together, so the jet pulls tend to point toward each other. For a pair of selected $b$-jets with transverse momenta $p_T^{b_1, b_2}$, we define an effective ``pull angle" $\theta_{\rm eff}$ by using the pull angles $\theta_t^{b_1,b_2}$ which the pulls of the two $b$-jets make with the chord joining the two jets in the $y-\varphi$ plane: $\theta_{\rm eff} = \left( ( \theta_t^{b_1} / \sigma(p_T^{b_1}) )^{2} + ( \theta_t^{b_2} / \sigma(p_T^{b_2}) )^{2} \right)^{1/2}$. Here $\sigma(p_T^{b_i}) = a p_T^{b_{i}} + b$ reflects the jet $p_T$ dependence of the standard deviation of the pull angle for the two $b$-jets produced from Higgs decay, as noticed in~\cite{Gallicchio:2010sw}. For the kinematic regime that we are considering, $a=-1$ TeV$^{-1}$ and $b=1.5$ are assumed.

If the $b$-jet pairs in each event are ranked according to these variables, noticeable differences emerge. Figure~\ref{rank} shows the $p_T$ vs. $\theta_{\rm eff}$ ranks of correct and wrong pairs in Case I at the 14 TeV LHC. The distribution of pairs in the ranking plane can be used to improve the Higgs search sensitivity.

\begin{table}
\begin{center}

\begin{tabular}{|c|c|c||c|c|}
\hline
$\sqrt{s}=14$ TeV &
    $t\bar{t}$+jets     &$t\bar{t}b\bar{b}$& Case I & Case II\\ \hline
Events & $5.2\times10^7$ & $8.2\times10^5$ & 26176  & 822275   \\ \hline
Cut 1 & $3.5\times10^7$  & 474234          & 20600  & 406296   \\ \hline
Cut 2 & 88700            & 12077           & 961    & 790   \\ \hline
Cut 3 & 51 / 79          & 442 / 796       & 567    & 411   \\ \hline
Cut 4 & 29 / 23          & 351 / 366       & 547    & 361   \\ \hline
Choice A & 20 / 11       & 5+157 / 4+157   & 99+215 & 76+126   \\ \hline
Choice B & 20 / 12       & 4+166 / 4+159   & 91+219 & 95+116   \\ \hline
Choice C & 13 / 13       & 5+104 / 3+104   & 78+147 & 71+65   \\ \hline
Choice D & 19 / 22       & 2+189 / 4+159   & 89+322 & 68+239   \\ \hline
\end{tabular}

\end{center}
\caption{Cut flows for the benchmark point at 14 TeV. The cut flow of some SM events is labeled as `Case I / Case II', where different cuts are used at the same stage. $t\bar{t}b\bar{b}$ includes QCD, $t\bar{t}Z$ and $t\bar{t}h$. Choice A, B, C, D rows correspond to the $\theta_{\rm eff}$, $p_T$ and $p_T-\theta_{\rm eff}$ plane, and min$(|m_{bb}-m_h|)$ pair selection methods, respectively. Bins labeled as (N + M) show (Higgs + wrong) $b$-jet pairs with $100 {\rm \ GeV}<m_{bb}<140 {\rm \ GeV}$.}
\label{cutflow}
\end{table}

\begin{figure}[h]
\centering
\includegraphics[width=0.32\textwidth]{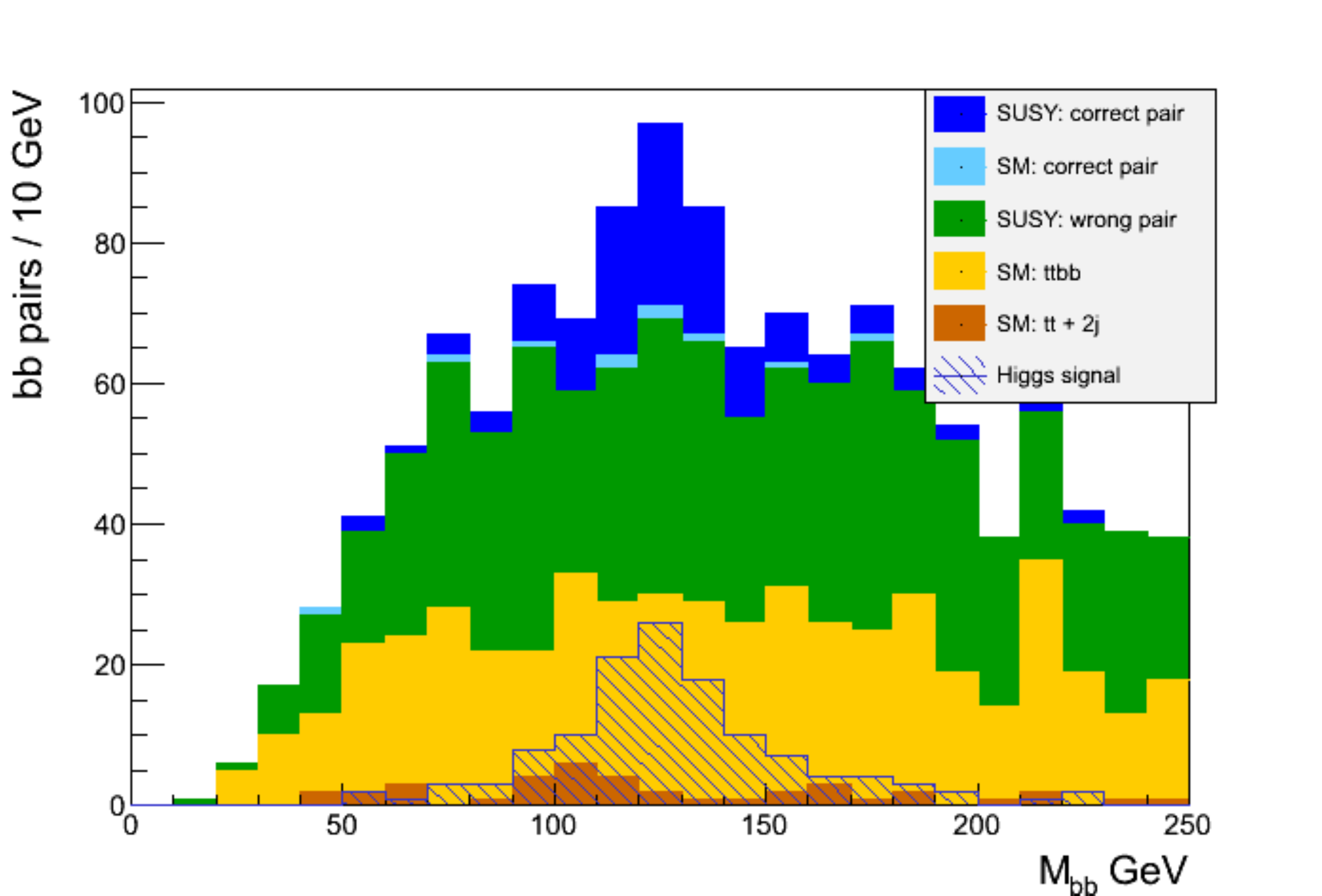}
\caption{B-jet pair invariant mass, Case I at 14 TeV. \label{invmass1}}
\end{figure}

\begin{figure}[!h]
\centering
\includegraphics[width=0.3\textwidth]{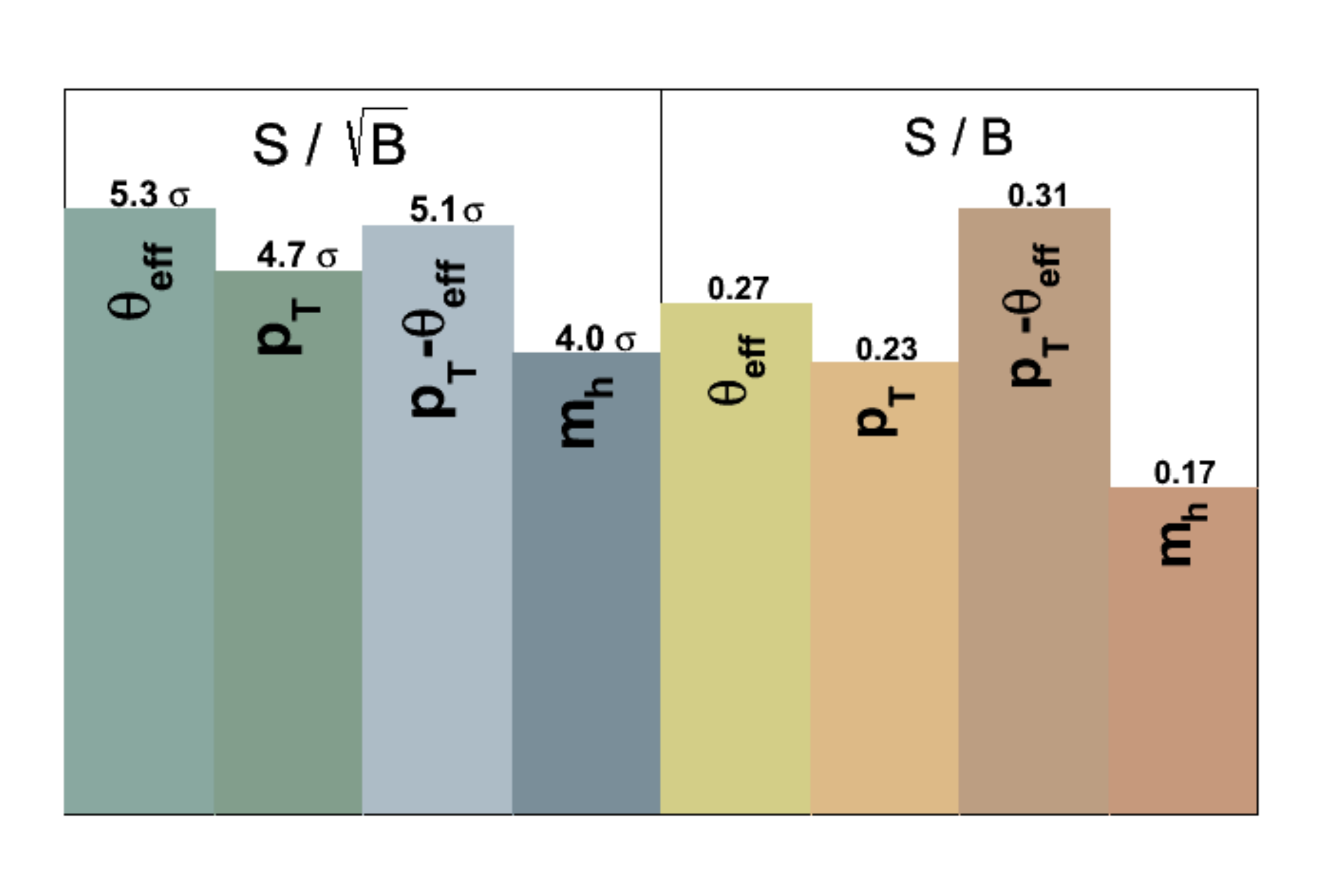}
\caption{Signal significance using various methods of choosing pairs, Case I at 14 TeV, with 40 fb$^{-1}$ of data.}
\label{significance1}
\end{figure}

\begin{figure}[h]
\centering
\includegraphics[width=0.32\textwidth]{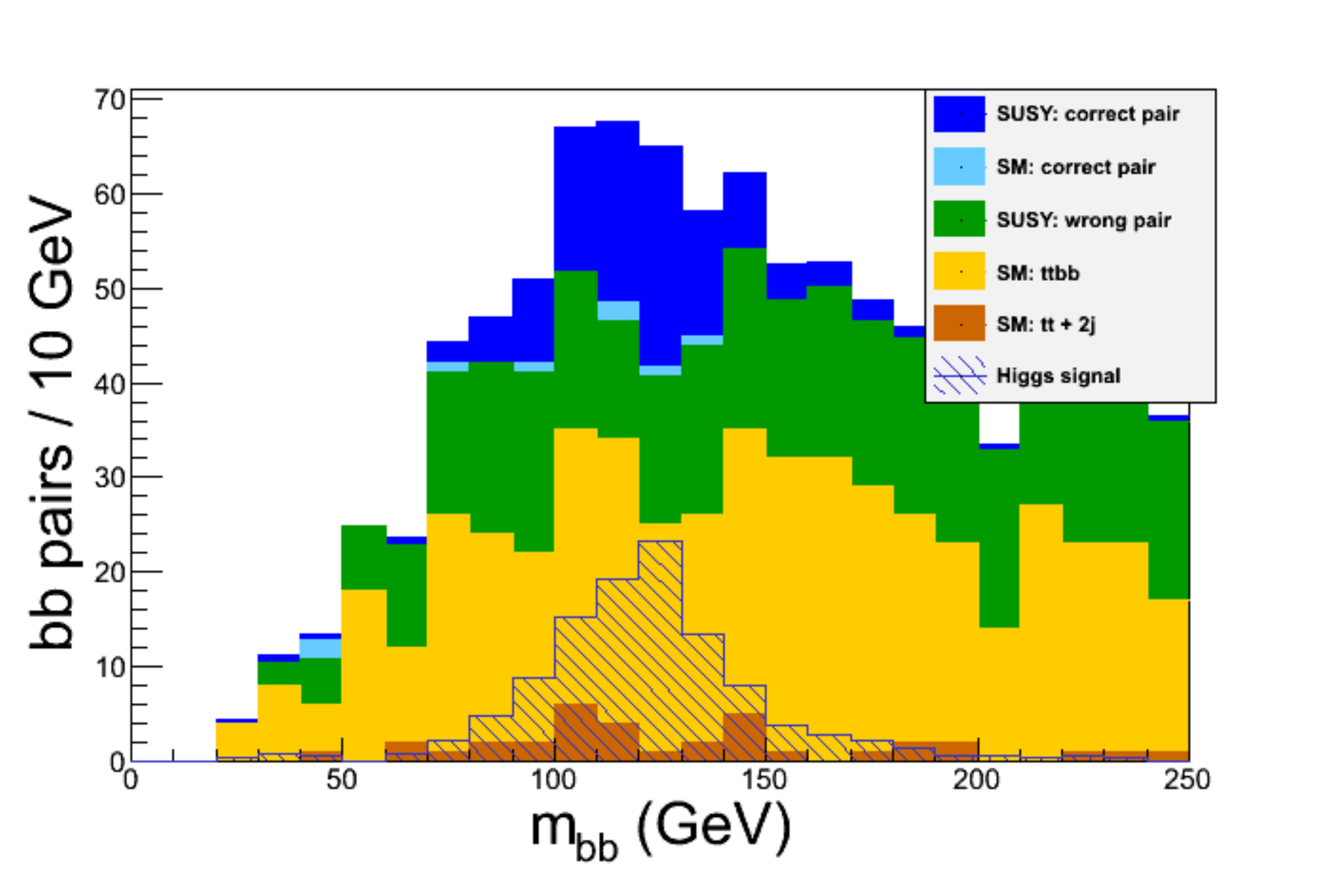}
\caption{B-jet pair invariant mass, Case II at 14 TeV. \label{invmass2}}
\end{figure}

\begin{figure}[!h]
\centering
\includegraphics[width=0.3\textwidth]{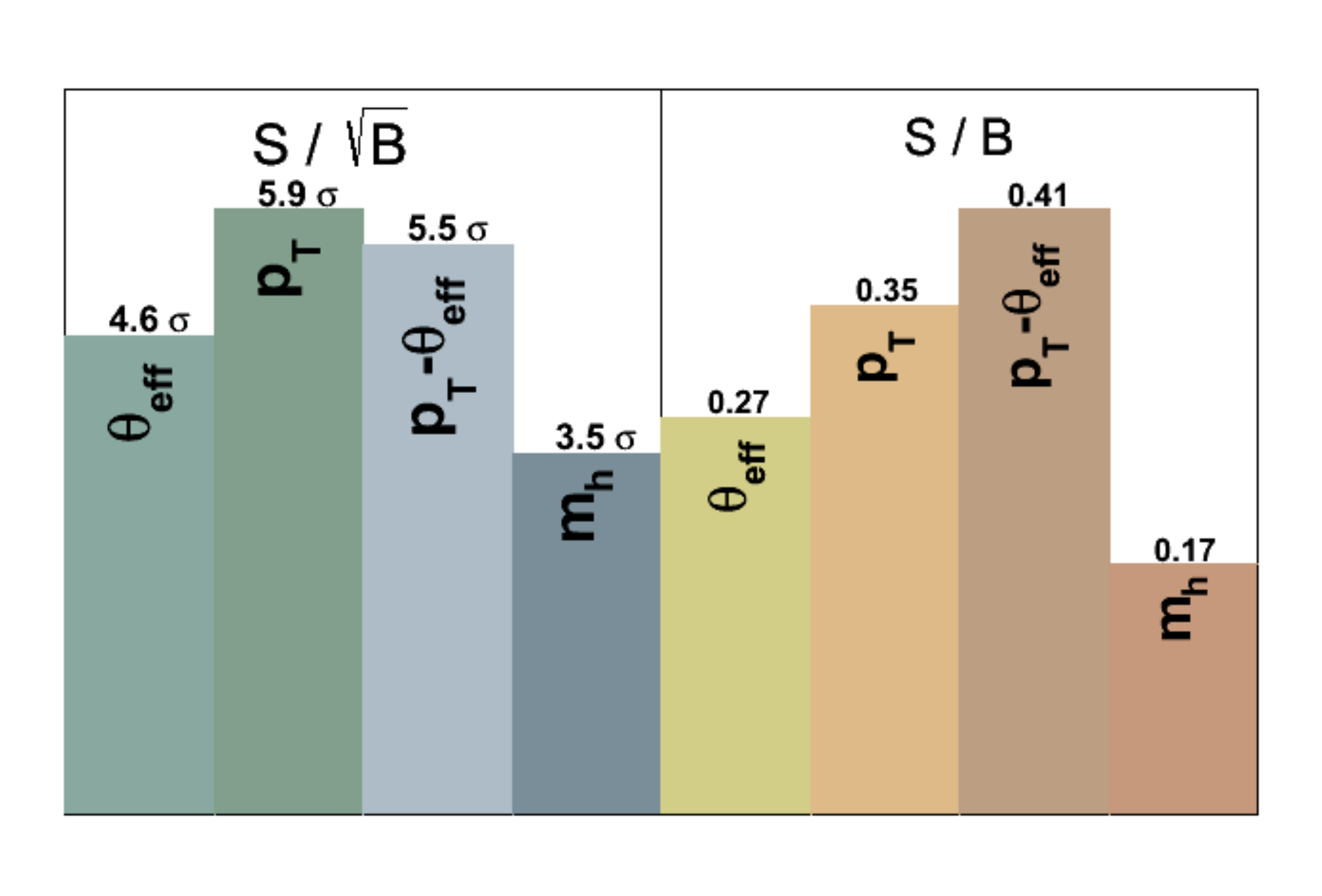}
\caption{Signal significance using various methods of choosing pairs, Case II at 14 TeV, with 40 fb$^{-1}$ of data.}
\label{significance2}
\end{figure}

Complete cut flows of both benchmark points at 14 TeV are presented in table~\ref{cutflow}. The number of events have been rescaled to the cross section of each process. A first observation is a relatively large sensitivity to our chosen benchmark points. From Table~\ref{cutflow} the sensitivities in both cases are $\gtrsim 20 \sigma$ for 40fb$^{-1}$ data at the 14 TeV LHC. Again, note that the contribution labeled ``SUSY: wrong pair" in these figures is a `background' only for the Higgs search, but indicates a large signal for the discovery of SUSY itself. To combine the $p_T$ and $\theta_{\rm eff}$ ranking strategies, we require that the selected pairs fall within the triangular region where rank($p_T$) + rank ($\theta_{\rm eff}$) $\leq$ 5. This strategy is used for the invariant mass plots of the selected $b$-jet pairs in Figure~\ref{invmass1} and Figure~\ref{invmass2}, for the SM-like Higgs candidates. The local significances for different strategies of the resonance reconstruction are shown in Figure~\ref{significance1} and Figure~\ref{significance2}. For the best option, the correctly paired Higgs $b$-jets give a local significance of $> 5\sigma$ in both cases for 40 fb$^{-1}$ of data, with the local mass window taken to be 100 GeV to 140 GeV. Though rank($p_T$) + rank ($\theta_{\rm eff}$) $\leq$ 5 does not increase the values of $S/\sqrt{B}$ significantly, it improves the $S/B$ to be above 0.3 and 0.4 for the two cases, respectively. This can potentially decrease the impact of systematic uncertainties. Each of these strategies is superior to naively selecting the $b$ pair with invariant mass closest to 125 GeV; for both benchmark points, this naive choice gives a sensitivity of $\sim4\sigma$ in Case I and $\sim3.5\sigma$ in Case II, with $S/B\sim0.17$ in both cases.

These results may be improved. Our $b$-tagging efficiency is rather conservative~\cite{:ATLAS-CONF-2011-102}. If the b-tagging efficiency is assumed to be 0.7, similar to the approach in~\cite{Plehn:2009rk,:ATL-PHYS-PUB-2009-088}, then the significances for various strategies are expected to be universally increased by a factor $\sim 1.5$.

Although the foregoing discussion is confined within the supersymmetric scenarios with relatively light top partners - its impact is fairly profound. It provides an independent way to discover the SM-like Higgs boson and to understand the mass origin of the SM fermions. In turn, searching for the Higgs boson using this strategy provides a direct way to test SUSY. In addition, the potential applicable scope of this strategy is broad. The first class of examples are some non-SUSY theories, such as little Higgs models~\cite{ArkaniHamed:2001nc}, composite Higgs models~\cite{Contino:2003ve}, and Randall-Sundrum models~\cite{Randall:1999ee}, where fermionic top partners $t'$ with a mass below 1 TeV are typically predicted~\cite{Matsedonskyi:2012ym}, given a 125 GeV SM-like Higgs boson, and the Higgs boson can be produced via $t' \to h t$, a way similar to the second mechanism discussed above. Actually, multiple $b$-tagging has been noticed to be useful in this context~\cite{AguilarSaavedra:2006gw}, while the Higgs reconstruction was thought to be a big problem due to the combinatorial background~\cite{Cacciapaglia:2011fx}.  Another class of examples are the SM Higgs search via the $t\bar th$ production and the non-standard Higgs ($H, A$) search via the $b\bar bH$ or $b\bar b A$ production in the MSSM. Though the involved kinematics are different, given $h, H, A \to b\bar b$, these analyses are expected to share the feature of suppressed $4j$ and $t\bar t, b\bar b + 2j$ background, leaving the continuum $t\bar t b\bar b$ or $b\bar b b\bar b$ as the main one. Then the Higgs resonance can be reconstructed with color-flow or kinematic variables. We leave the consideration of these interesting possibilities to future work.

\vspace*{-5pt}

\begin{center}
{\bf Acknowledgements}
\end{center}

\vspace*{-5pt}

We would like to thank T. Danielson, J. Gallicchio, D. Krohn, A. Menon, D. Morrissey, J. Richman, M. Schwartz, J. Shelton, J. Shu, S. Su, N. Toro, and L.T. Wang for useful discussions. Work is supported in part by DOE under grant DE-FG02-91ER40618.

\vspace*{-15pt}


\end{document}